\documentclass[prb,preprint]{revtex4-1} 

\usepackage{amsmath} 

\usepackage{graphicx} 
\usepackage{hyperref}
\usepackage{xcolor}

\newcommand{\be}{\begin{equation}}
\newcommand{\ee}{\end{equation}}
\newcommand{\bi}{\begin{itemize}}
\newcommand{\ei}{\end{itemize}}
\newcommand{\bea}{\begin{eqnarray}}
\newcommand{\eea}{\end{eqnarray}}

\newcommand{\p}{\partial}

\newcommand{\void}[1]{}

\begin{document}

\title{
Obtaining Maxwell's equations heuristically}
\author{Gerhard Diener$^{\dagger}$, J\"urgen Weissbarth, Frank Grossmann and R\"udiger Schmidt}
\affiliation{Technische Universit\"at Dresden, Institut f\"ur Theoretische Physik,
D-01062 Dresden, Germany}
\email{frank@physik.tu-dresden.de} 
\date{\today}

\begin{abstract}
Starting from the experimental fact that a moving charge
experiences the Lorentz force and applying the fundamental principles of simplicity 
(first order derivatives only) and linearity (superposition principle), we show that 
the structure of the microscopic Maxwell equations for the electromagnetic fields can be deduced 
heuristically by using 
the transformation properties of the fields under space inversion and time reversal.
Using the experimental facts of charge conservation and that electromagnetic
waves propagate with the speed of light together with Galileo invariance of the
Lorentz force allows us to introduce arbitrary electrodynamic
units naturally.
\end{abstract}

\maketitle

\section{Introduction}

Teaching an electrodynamics course one is faced with the question of whether
one should postulate Maxwell's equations, as one postulates
Newton's laws in an introductory classical mechanics course, or whether they
should be justified from experimental experience (Coulomb's law, Faraday's
induction law, Amp{\`e}re's law and the nonexistence of magnetic monopoles).
Depending on the choice made, the didactic approach is then either deductive 
(axiomatic) or inductive.

Here we give a heuristic derivation of the microscopic Maxwell equations. This 
derivation is based on the principles of simplicity (lowest order in space and time
derivatives), linearity (superposition principle) 
and the transformation properties of the fields under space inversion, 
$\vec{r}\to -\vec{r}$, and time reversal, $t\to -t$. The starting point of 
the derivation is the experimental fact that a (moving) charge feels the Lorentz force.
In addition, in order to define electrodynamic units, we use the experimental evidence 
of charge conservation, 
the fact that electromagnetic waves propagate at the speed of light $c$ and the
requirement of Galileo invariance of the Lorentz force for low velocities.  

Several attempts to deduce (``derive'') Maxwell's equations have been 
published.\cite{Kobe,Dyson,Heras1,Heras2} The approach that will be presented below is 
alternative because it does not need another dynamical equation, for example, 
the time-dependent Schr\"odinger equation \cite{Kobe} or Newton's law \cite{Dyson} 
as a starting point. Our derivation may serve as valuable 
background information for the lecturer in both approaches to teaching electrodynamics. 
It can be used as an a posteriori justification 
after the Maxwell equations have been postulated and/or obtained from pure experimental 
experience.

The presentation is structured as follows: in Sec.\ II we briefly review the
classification of vectors and scalars by their behavior under space inversion and time reversal.
In Sec.\ III we present the deduction of Maxwell's equations from the principles
of simplicity and linearity, with five undetermined multiplicative constants. 
Three of the five constants are determined in Sec.\ IV from well-established
experimental facts together with the demand of Galileo invariance of the Lorentz force. 
Fixing the final two constants leads to the natural introduction of 
three commonly used systems of units in electrodynamics.

\section{Polar and axial vectors and time reversal properties}

In a standard physics curriculum starting with classical mechanics,
the necessity of studying electrodynamics can be motivated by the implications of 
the particle property charge, be it fixed or moving. 
Experimental evidence shows that the Lorentz force on a test particle with
electric charge $q$, moving with velocity $\vec{v}$ is given by
\be\label{LK}
 \vec{F} = q\,(\vec{E}+\eta\,\vec{v}\times\vec{B}\,).
\ee
This equation postulates the existence of a local electric field $\vec{E}(\vec{r},t)$ and 
a local magnetic
induction $\vec{B}(\vec{r},t)$ at the position of the particle $\vec{r}$, generated by all
charge carrying, field generating particles except the test particle itself. 
A particle with the property magnetic charge is not known up to now. Thus the fields
are generated only by electric charges with charge density $\rho$ and moving
charges with electric current density $\vec{j}$ (see below).

Due to the multiplicative connection of charge and fields in Eq.\ (\ref{LK}),
their units can be chosen arbitrarily. E.\ g., an arbitrarily defined charge
unit $[q]$ fixes the dimension of the electric field $[E]$ and that of
the product $[\eta B]$ with again an arbitrary constant $\eta$ which finally
defines $[B]$ and thus fixes the relative dimension of both fields. 

Of central importance for the following is the fact that 
the vector character with respect to space inversion, $\vec{r}\to -\vec{r}$, is 
twofold:\cite{Emde,Jackson,Norb90}

\begin{figure}[htb]
\begin{center}
\includegraphics[width=0.65\linewidth]{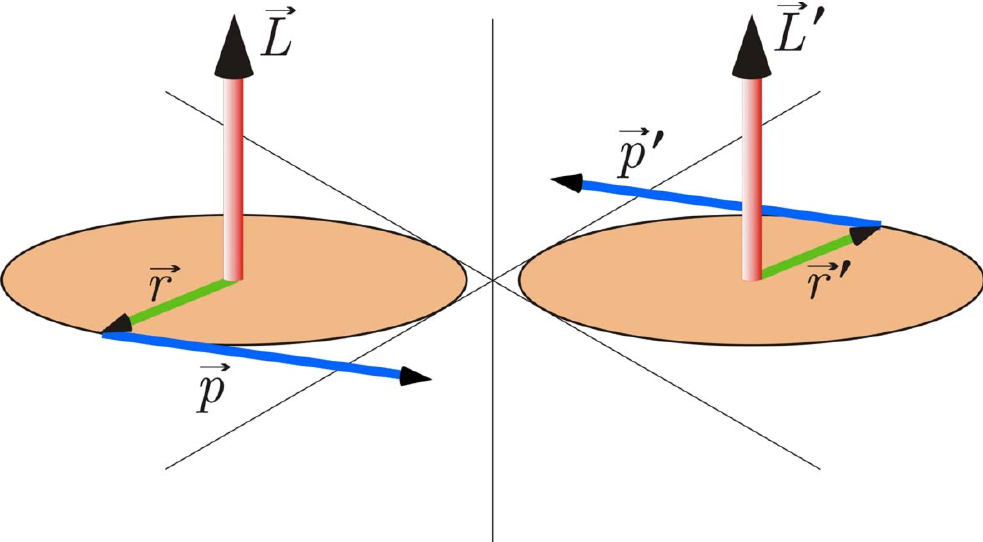}
\caption{An example for the transformation properties of polar vectors (position $\vec{r}$ and
momentum $\vec{p}$) and an axial vector (angular momentum $\vec{L}=\vec{r}\times\vec{p}$) under 
space inversion
\cite{wiki}.}
\label{fig:vec}
\end{center}
\end{figure}

\bi
\item
Vectors $\vec{p}$, that transform according to
\be
\vec{p}\stackrel{\vec{r}\to -\vec{r}}{\longrightarrow}\vec{p}~'=-\vec{p},
\ee
under inversion are called polar vectors (or just vectors). 
\item
Vectors $\vec{a}$, that transform according to
\be
\vec{a}\stackrel{\vec{r}\to -\vec{r}}{\longrightarrow}\vec{a}~'=\vec{a},
\ee
under inversion are called axial vectors (or pseudo vectors). They need a (right) hand rule for their definition.
\ei
As an example from classical mechanics, we explicitly depict the
transformation properties of position $\vec{r}$, momentum $\vec{p}$ and angular momentum $\vec{L}$ in 
Fig.\ \ref{fig:vec}.

For the multiplication of different types of vectors the following rules hold:
\bea
\vec{p}_1 \times \vec{p}_2 &=& \vec{a}
\\
\vec{a}_1 \times \vec{a}_2 &=& \vec{a}
\\
\label{eq:akreuzp}
\vec{a}_1 \times \vec{p}_1 &=& \vec{p}.
\eea

Furthermore, in the same manner as there are vectors and pseudo vectors, also
scalars $S$ transforming according to
\be
S\stackrel{\vec{r}\to -\vec{r}}{\longrightarrow}S'=S
\ee
under inversion and pseudo scalars $S^\ast$ with 
\be
S^\ast\stackrel{\vec{r}\to -\vec{r}}{\longrightarrow}{S^{\ast}}'=-S^\ast
\ee
can be defined. They can be generated, e.\ g., by taking the following products
\bea
\vec{p}_1 \cdot \vec{p}_2 &=& S
\\
\vec{a}_1 \cdot \vec{a}_2 &=& S
\\
\vec{p}_1 \cdot \vec{a}_1 &=& S^\ast
\label{eq:ps3}
\eea

We can now identify the vector, respectively scalar character of the quantities appearing in the 
Lorentz force, together with their behavior under the reversal of time.
\begin{enumerate}
\item[(i)]
Because the mass $m$ is a scalar, $\vec{F}=m\ddot{\vec{r}}$ is a {\em polar} vector with $\vec{F}(-t)=\vec{F}(t)$.
Therefore, because $q$ in (\ref{LK}) is a scalar, $\vec{E}$ is a {\em polar} vector 
with $\vec{E}(\vec{r},-t)=\vec{E}(\vec{r},t)$.
\item[(ii)]
Because the velocity $\vec{v}$ is a {\em polar} vector with $\vec{v}(-t)=-\vec{v}(t)$,
due to the vector product in (\ref{LK}),
$\vec{B}$ is an {\em axial} vector with $\vec{B}(\vec{r},-t)=-\vec{B}(\vec{r},t)$.
\end{enumerate}
This unequal nature of the two field vectors of electrodynamics has been known and used
in teaching for a long time.\cite{foot1} An early appearance and critical discussion in the 
literature can be found in.\cite{Emde}

For future reference we state that the field generating charge density is defined by
$\varrho:=\lim_{\Delta V\to 0}\Delta Q/\Delta V$, where $\Delta V>0$ because of the discrete nature
of charge. Here, as usual, it is idealized as a continuous scalar field
with $\varrho(\vec{r},t)=\varrho(-\vec{r},t)$ and $\varrho(\vec{r},-t)=\varrho(\vec{r},t)$ and thus the 
field generating current density 
\be\label{curr}
\vec{j}(\vec{r},t)=\varrho(\vec{r},t)~\vec{\nu}(\vec{r},t),
\ee
with the velocity field $\vec{\nu}(\vec{r},t)$ of the charge density (not to be confused with the
test particle velocity $\vec{v}$ in Eq.\ (\ref{LK})), is a {\em polar} vector 
with $\vec{j}(\vec{r},t)=-\vec{j}(-\vec{r},t)$ and $\vec{j}(\vec{r},-t)=-\vec{j}(\vec{r},t)$.

\section{Generating the Maxwell equations: The basic idea}

The classic text book by Jackson \cite{Jackson} contains a discussion showing that the
Maxwell equations do indeed equate quantities with the same transformation
behavior under time reversal and space inversion. 
The idea to be presented here is that this behavior together with 
the Lorentz force and the heuristic demand for simplicity and linearity allows to {\it deduce} the 
structure of the Maxwell equations. The principles  of simplicity and linearity 
together with the symmetry properties of the fields have already been employed by Migdal \cite{Migdal} 
to derive the homogeneous curl equations. 
The program that we will follow here is to equate field generating quantities $\varrho$ and $\vec{j}$ 
with derivatives of the fields with respect to time or space up to at most first order, and, 
so that the fields and the inhomogeneous terms fulfill the superposition principle.\cite{foot4}

Firstly, as discussed above, there is no experimental evidence of 
magnetic charges. Thus there do not exist any field generating pseudo scalars and the only
pseudo scalar (\ref{eq:ps3}) that can be generated by taking first order spatial derivatives of any of the
fields, which is $\vec{\nabla}\cdot\vec{B}$, has to vanish and we find
\be\label{divB}
\vec{\nabla}\cdot\vec{B}=0.
\ee
Nevertheless, magnetic monopoles are a topic of current research, especially because,
based on an idea by Dirac, they may explain the discrete nature of the electric 
charge.\cite{Schwinger} Furthermore, they may be helpful as a didactic tool.\cite{Crawford}
We note in passing that there is a very simple physical argument for the vanishing
divergence of $\vec{B}$: if this divergence would be non-zero, a magnetic induction field $\vec{B}$
proportional to $\vec{r}$ would exist (as is well-known from the electric field case)
and the Lorentz force (\ref{LK}) would contain terms proportional
to $\dot{\vec{r}}\times\vec{r}$, i.\ e., terms proportional to the angular momentum that
would lead to out of plane acceleration. For a charged particle corresponding trajectories 
have, however, never been observed in pure magnetic induction fields. 
Concluding the discussion of (\ref{divB}), we mention that mixed terms of the correct symmetry, 
as, e.\ g., $\vec{j}\cdot\dot{\vec{B}}$ in (\ref{divB}) would violate the 
superposition principle\cite{foot4}. 

Secondly, due to the fact that $\dot{\vec{B}}$,
as well as according to (\ref{eq:akreuzp}) also $\vec{\nabla}\times\vec{E}$,
are the only axial vectors without sign change on time reversal they have to appear in
one equation together. A field generating axial vector does not exist 
and the heuristic demand for simplicity and linearity lead us to the equation
\be\label{rotE}
\vec{\nabla}\times\vec{E} + \chi\,\dot{\vec{B}}=0
\ee
with an undetermined constant $\chi$.

Thirdly, the only scalar that can be generated from first derivatives of the fields without 
sign change on time reversal is $\vec{\nabla}\cdot\vec{E}$. This quantity must be 
determined by the only scalar source term $\varrho$ and therefore we can write
\be\label{divE}
\vec{\nabla}\cdot\vec{E}=\alpha\, \varrho
\ee
with a yet undetermined constant $\alpha$. Again we note that mixed terms of the correct symmetry, 
as, e.\ g., $\vec{j}\cdot\dot{\vec{E}}$ in (\ref{divE}) would violate the superposition 
principle\cite{foot4}, whereas a term proportional to $\ddot{\rho}$ violates the
use of maximally first order derivatives.

Finally, the only polar vectors with sign change on time reversal that can be
generated from the fields by first order derivatives are $\dot{\vec{E}}$ and 
$\vec{\nabla}\times\vec{B}$; they therefore appear in one equation together with 
$\vec{j}$, that reads
\be\label{rotB}
\vec{\nabla}\times\vec{B}+\kappa\, \dot{\vec{E}}=\beta\, \vec{j},
\ee
with the two constants $\kappa,\beta$ still to be determined.

The homogeneous (\ref{divB},\ref{rotE}) as well as the inhomogeneous 
Maxwell equations (\ref{divE},\ref{rotB}) constructed in
this way still contain 4 undetermined constants $\chi,\alpha,\kappa$, and $\beta$.
Together with the constant $\eta$ appearing in the Lorentz force, Eq. (\ref{LK}), 5 constants
still have to be determined. 

\section{Determining the constants and choice of units}

As an experimental fact, the field generating quantities $\rho$ and $\vec{j}$ have to fulfill the 
equation of continuity. In addition we know that the Lorentz force has to be invariant under
Galileo transformation for small $v\ll c$ with the speed of light $c$ (Lorentz invariance 
adds corrections proportional to $(v/c)^2$). Furthermore, from experimental evidence as well as
from the theory of special relativity, we know that the universal propagation
speed $c$ has to be contained in the wave equation, which can be deduced from the equations obtained in Sec.\ III
(see below). For the five constants we therefore have 3 equations, meaning 
that there is a degree of ambiguity, which will be resolved by the discussion of
three possible, frequently used systems of units, the ``Syst{\`e}me International d' unit{\'e}s'' 
(SI units), the Gaussian system (cgs units), and the Heaviside-Lorentz units.

\subsection{Charge conservation}

Taking the divergence of (\ref{rotB}) together with (\ref{divE}) leads to
\be
\beta \vec{\nabla}\cdot\vec{j} = \kappa\,\vec{\nabla}\cdot\dot{\vec{E}}= \kappa\, \alpha\,\dot{\varrho}.
\ee
Thus the fundamental equation of continuity $\dot{\varrho}+\vec{\nabla}\cdot\vec{j}=0$ is fulfilled if
\be
\kappa\alpha=-\beta.
\label{eq:alpha}
\ee

\subsection{Invariance of the Lorentz force}

The Lorentz invariance of the Lorentz force for small velocities, $v\ll c$, in leading
order means Galileo invariance.  For the purposes of determining the
constant $\chi$ it is therefore sufficient to consider the  Galileo transformation 
$\;\vec{r}\to\vec{r}\,'=\vec{r}-\vec{v}_0t$; $\;\;\vec{v}\,'=\vec{v}-\vec{v}_0$.  
Applied to (\ref{LK}) this leads to
\be
\vec{F}=q(\vec{E}+\eta\vec{v}\times\vec{B}\,)
 = q(\vec{E}+\eta\vec{v}_0\times\vec{B} + \eta\vec{v}\,'\times\vec{B}\,)
 \stackrel{!}{=} q(\vec{E}\,' + \eta\vec{v}\,'\times\vec{B}\,').
\ee
For the fields, the postulated invariance of the Lorentz force means
\be\label{GTF}
\vec{E}\,' = \vec{E}+\eta\vec{v}_0\times\vec{B}\,,\quad\vec{B}\,'=\vec{B}.
\ee
Using a calculation similar to that in,\cite{Jammer} we find on the other hand
\bea
 \frac{\p}{\p\vec{r}}\times\vec{E}(\vec{r},t) &=&
        \frac{\p}{\p\vec{r}\,'}\times\vec{E}(\vec{r}\,'+\vec{v}_0t,t) 
 \\
 \left.\frac{\p\vec{B}(\vec{r},t)}{\p t}\right|_{\vec{r}=\mbox{\tiny const}} \hspace*{-2em} &=&
 \left.\frac{\p\vec{B}(\vec{r}\,'+\vec{v}_0t,t)}{\p t}\right|_{\vec{r}\,'=\mbox{\tiny const}}
 \hspace*{-2em} -\left(\vec{v}_0\cdot\frac{\p}{\p\vec{r}\,'}\right)\vec{B}
 \\
  &=& \left.\frac{\p\vec{B}(\vec{r}\,'+\vec{v}_0t,t)}{\p t}\right|_{\vec{r}\,'}
  + \frac{\p}{\p\vec{r}\,'}\times\left(\vec{v}_0\times\vec{B}\,\right).
\eea
For (\ref{rotE}) this leads to
\be
\frac{\p}{\p\vec{r}}\times\vec{E}+\chi\dot{\vec{B}} =
 \frac{\p}{\p\vec{r}\,'}\times\left(\vec{E}+\chi\vec{v}_0\times\vec{B} \,\right)
        + \chi\left.\frac{\p\vec{B}}{\p t}\right|_{\vec{r}\,'} \stackrel{!}{=}
 \frac{\p}{\p\vec{r}\,'}\times\vec{E}\,'+\chi\frac{\p\vec{B}\,'}{\p t}=0.
\ee
Therefore 
\be
\vec{E}\,' = \vec{E}+\chi\vec{v}_0\times\vec{B}
\ee
has to hold and by comparison with (\ref{GTF}) we find 
\be
\label{eq:beta}
\chi = \eta.
\ee

\subsection{Wave equation}

The wave equation for vanishing inhomogeneities $\varrho$ and $\vec{j}$ follows by taking 
the curl of (\ref{rotE}) and using (\ref{divE}) and (\ref{rotB}). With
$\vec{\nabla}\times(\vec{\nabla}\times\vec{E})=\vec{\nabla}(\vec{\nabla}\cdot\vec{E}\,) - \Delta \vec{E}$
this leads to
\be
\vec{\nabla}(\vec{\nabla}\cdot\vec{E}\,) - \Delta \vec{E}
        + \chi\,\vec{\nabla}\times\dot{\vec{B}}
=- \Delta \vec{E}-\chi\kappa\,\ddot{\vec{E}}=0.
\ee
Finally, using (\ref{eq:beta}) we find that 
\be
\label{eq:const}
\eta\kappa = -\frac{1}{c^2}
\ee
has to hold for the remaining constants with a velocity $c$, determined by measuring the
velocity of electromagnetic waves in vacuum, which turns out to be $c=$2.9986$\times 10^8$m/s,
the (universal) vacuum speed of light.
It is worthwhile to note that a velocity with the same numerical value can be determined by pure
electrostatic and magnetostatic measurements (``$c$ equivalence principle'').\cite{Heras3, Heras4}

\subsection{Final choice of the system of units}

The constant $\chi$ is now fixed by (\ref{eq:beta}) and for 
$\kappa,\alpha,\beta$, and $\eta$ we have the two relations
(\ref{eq:alpha}) and (\ref{eq:const}), leading to $\alpha/(\eta\beta)=c^2$.\cite{Heras1} 
Two of the constants can be chosen arbitrarily. Common choices are:
\begin{itemize}
\item[(i)]SI units: $\eta=1$ leading to $\kappa=-1/c^2$, and $\beta=4\pi\times 10^{-7}$N/A$^2\equiv \mu_0$ leading to  
$\alpha=\mu_0c^2\equiv 1/\varepsilon_0$ and thus to the Lorentz force 
\be
\vec{F}=q[\vec{E}(\vec{r},t)+\vec{v}\times\vec{B}(\vec{r},t)],
\ee
and the rationalized Maxwell equations  (without explicit appearance of $\pi$)
\bea
\vec{\nabla}\times\vec{E}(\vec{r},t) &=& - \dot{\vec{B}}(\vec{r},t) 
\\
\vec{\nabla}\times\vec{B}(\vec{r},t) &=& \mu_0 \vec{j}(\vec{r},t) + \mu_0 \varepsilon_0 \dot{\vec{E}}(\vec{r},t) 
\\
\vec{\nabla}\cdot \vec{E}(\vec{r},t) &=& \frac{\varrho(\vec{r},t)}{\varepsilon_0} 
\\
\vec{\nabla}\cdot \vec{B}(\vec{r},t) &=& 0.
\eea
\item[(ii)]cgs units: $\eta=1/c$ leading to $\kappa=-1/c$, and $\beta=4\pi/c$ leading to $\alpha=4\pi$, and 
the Lorentz force 
\be
\label{LK2}
\vec{F}=q[\vec{E}(\vec{r},t)+\frac{\vec{v}}{c}\times\vec{B}(\vec{r},t)],
\ee
where the electric field and the magnetic induction have the same units. The Maxwell equations read
\bea
\vec{\nabla}\times\vec{E}(\vec{r},t) &=& - \frac{1}{c}\dot{\vec{B}}(\vec{r},t)
\\
\vec{\nabla}\times\vec{B}(\vec{r},t) &=& \frac{4\pi}{c} \vec{j}(\vec{r},t) + \frac{1}{c} \dot{\vec{E}}(\vec{r},t) 
\\
\vec{\nabla}\cdot\vec{E}(\vec{r},t) &=& 4\pi \varrho(\vec{r},t)
\\
\vec{\nabla}\cdot\vec{B}(\vec{r},t) &=& 0.
\eea
\item[(iii)]Heaviside-Lorentz units: $\eta=1/c$ leading to $\kappa=-1/c$, and $\beta=1/c$ leading to  $\alpha=1$. This gives 
again the Lorentz force (\ref{LK2}) from above but with different field units (although electric field $\vec{E}$ and magnetic
induction $\vec{B}$ have again the same units) and leads to rationalized Maxwell equations, but without the explicit 
appearance of $\varepsilon_0$ and $\mu_0$
\bea
\vec{\nabla}\times\vec{E}(\vec{r},t) &=& - \frac{1}{c}\dot{\vec{B}}(\vec{r},t)
\\
\vec{\nabla}\times\vec{B}(\vec{r},t) &=& \frac{1}{c} \vec{j}(\vec{r},t) + \frac{1}{c} \dot{\vec{E}}(\vec{r},t) 
\\
\vec{\nabla}\cdot\vec{E}(\vec{r},t) &=& \varrho(\vec{r},t)
\\
\vec{\nabla}\cdot\vec{B}(\vec{r},t) &=& 0.
\eea
\end{itemize}
In this way the units are introduced naturally by fixing two remaining constants appearing in 
the Maxwell equations and the Lorentz force.

\section{Summary}

Starting from the Lorentz force as an experimental fact and using the principles of simplicity and 
linearity, we could show that the structure of the Maxwell equations
for the electromagnetic fields in vacuum can be deduced. These are two (pseudo)scalar equations for the
divergence of the respective fields and two (pseudo)vector equations for the curl of the fields, in accord with the
fundamental theorem of vector calculus. The vector 
character (polar or axial) of the fields and their transformation properties under time reversal 
are at the heart of our presented approach. 
Charge conservation, the Galileo invariance of the Lorentz force and the fact that
electromagnetic waves propagate at the speed of light enabled us to
fix three of the five undetermined constants in the Lorentz force and the 4 deduced Maxwell equations.
As an additional useful result of the calculation,
arbitrary systems of units can be introduced, which is didactically more pleasing than to
postulate the units a priori.

\begin{acknowledgments}
Helpful comments on the manuscript by Larry Schulman and valuable
discussions with Klaus Becker are gratefully acknowledged.
\end{acknowledgments}

\vspace{1cm}
\noindent
$^\dagger$ deceased on August 17, 2009


\begin{thebibliography}{5}
\bibitem{foot1} The different nature of $\vec{E}$ and $\vec{B}$ is apparent also
in the construction of the electromagnetic field tensor in the covariant formulation of electrodynamics.
\bibitem{foot4} The superposition principle for an inhomogeneous, partial differential equation, with linear
differential operator $L$ states that with two solutions obeying $Lu_1=g_1$ and $Lu_2=g_2$ also the sum $u_1+u_2$ 
is a solution for the inhomogeneity $g_1+g_2$.
\end{thebibliography}
\end{document}